\documentclass[useAMS,usenatbib]{mn2e} 
\usepackage{graphicx} 
\title{A possible signature of annihilating dark matter} 
\author[Chan]{Man Ho Chan \thanks{chanmh@eduhk.hk}
\\ Department of Science and Environmental Studies, The Education University of Hong Kong, Tai Po, Hong Kong}

\begin{document}

\date{Accepted XXXX, Received XXXX}

\pagerange{\pageref{firstpage}--\pageref{lastpage}} \pubyear{XXXX}

\maketitle

\label{firstpage}

\date{\today}

\begin{abstract}
In this article, we report a new signature of dark matter annihilation based on the radio continuum data of NGC 1569 galaxy detected in the past few decades. After eliminating the thermal contribution of the radio signal, an abrupt change in the spectral index is shown in the radio spectrum. Previously, this signature was interpreted as an evidence of convective outflow of cosmic ray. However, we show that the cosmic ray contribution is not enough to account for the observed radio flux. We then discover that if dark matter annihilates via the 4-e channel with the thermal relic cross section, the electrons and positrons produced would emit a strong radio flux which can provide an excellent agreement with the observed signature. The best-fit dark matter mass is 25 GeV.
\end{abstract}


\section{Introduction}
In the past decade, the detections of high-energy positrons by HEAT \citep{Beatty}, PAMELA \citep{Adriani} and AMS-02 \citep{Aguilar,Accardo,Aguilar2} revealed some excess positron emissions in our galaxy. On the other hand, gamma-ray observations indicate some excess GeV gamma-rays emitted from our galactic center \citep{Abazajian,Calore,Daylan}. If dark matter annihilates, a large amount of positrons and gamma-ray photons would be produced. Therefore, many studies suggest that the excess positron and gamma-ray emissions can be explained by annihilating dark matter with mass $m \sim 10-100$ GeV \citep{Boudaud,Mauro,Abazajian,Calore,Daylan}. It is also surprising that the best-fit annihilation cross section is close to the thermal relic cross section $\sigma v=2.2 \times 10^{-26}$ cm$^3$ s$^{-1}$ \citep{Mauro,Daylan}, which is predicted in standard cosmology \citep{Steigman}. However, recently, gamma-ray observations of the Milky Way dwarf spheroidal satellite (MW dSphs) galaxies by Fermi-LAT put very tight constraints on dark matter mass and annihilation cross section \citep{Ackermann,Albert}. Furthermore, many studies suggest that pulsars' emission in our galaxy can account for the GeV gamma-ray and positron excess \citep{Hooper,Yuksel,Linden,Delahaye,Brandt,Bartels,Ajello}. Generally speaking, the most popular models of dark matter interpretation (e.g. annihilation via $b\bar{b}$ or $\mu^+\mu^-$) of the gamma-ray and positron excess are now disfavored \citep{Ajello}. 

In this article, we revisit the radio continuum data of NGC 1569 galaxy and perform a theoretical analysis with the annihilation model of dark matter. We show that the non-thermal radio spectrum of NGC 1569 exhibits a new possible signature of dark matter annihilation. Generally speaking, many theoretical models predict that dark matter particles can annihilate to give high-energy photons, electrons, positrons and neutrinos. For example, dark matter annihilation can first give a pair of electron and positron (the $e^+e^-$ channel) and then the electron-positron pair would generate a cascade of photons, electrons, positrons and neutrinos with different energies. Each annihilation channel can produce a unique electron energy spectrum. These high-energy electrons and positrons would generate synchrotron radiation (in radio frequencies) due to strong magnetic field in a galaxy. Therefore, if the radio signal detected mainly originates from the electrons and positrons produced from dark matter annihilation, we can probe the original electron and positron spectrum injected and infer the possible annihilation channel and dark matter rest mass. 

\section{Radio continuum data of NGC 1569}
NGC 1569 galaxy is a very good candidate for investigation because it is a nearby dark matter dominated dwarf galaxy (distance $=3.36 \pm 0.20$ Mpc) \citep{Johnson}. Also, the baryonic content is very small so that the radio flux contribution due to baryons is not significant. It has a relatively high magnetic field $B=14 \pm 3~ \mu$G (Local Group dwarf galaxies: $B=4.2 \pm 1.8~ \mu$G) \citep{Chyzy} so that the cooling timescale of the high-energy electrons and positrons is much smaller than their diffusion timescale (see the discussion below). In other words, most of the high-energy electrons and positrons would loss all of their energy via synchrotron cooling before traveling to a large distance and the radio flux emitted would be enhanced. 

The radio signals of NGC 1569 for different frequencies ($\nu=38$ MHz$-$24.5 GHz) were obtained in the past few decades \citep{Israel,Lisenfeld,Kepley}. The flux density is about $400$ mJy at 1.4 GHz (1 mJy$=10^{-29}$ W m$^{-2}$ Hz$^{-1}$). The overall spectral index for the radio flux density $S \propto \nu^{-\alpha}$ is $\alpha=0.47$ \citep{Lisenfeld,Kepley}. If we eliminate the thermal emission flux density $S_{\rm thermal}=100(\nu/\rm GHz)^{-0.1}$ mJy from the observed radio flux for different frequencies \citep{Lisenfeld}, we can obtain the non-thermal flux density. The calculated non-thermal radio flux density is shown in Table 1. We can see that the spectral index in the non-thermal radio spectrum has an abrupt break near $\nu=5$ GHz (see Fig.~1).

\begin{table}
\caption{The non-thermal radio flux density of NGC 1569 for different frequencies \citep{Lisenfeld}.}
 \label{table1}
 \begin{tabular}{@{}lc}
  \hline
  $\nu$ (GHz) &  Radio flux density (mJy) \\
  \hline
  0.038 & $2161\pm 580$ \\
  0.0575 & $1467 \pm 500$ \\
  0.151 & $819 \pm 180$ \\
  0.61 & $505 \pm 20$ \\
  1.415 & $323 \pm 35$ \\
  1.465 & $189 \pm 30$ \\
  1.49 & $315$ \\
  2.695 & $179 \pm 50$ \\
  2.7 & $274 \pm 23$ \\
  4.75 & $177 \pm 20$ \\
  4.85 & $117 \pm 19$ \\
  4.919 & $125 \pm 20$ \\
  4.995 & $195 \pm 20$ \\
  5 & $191 \pm 42$ \\
  6.63 & $152 \pm 30$ \\
  8.415 & $44 \pm 12$ \\
  10.7 & $81 \pm 20$ \\
  15.36 & $40 \pm 13$ \\
  24.5 & $23 \pm 8$ \\
  \hline
 \end{tabular}
\end{table}

\begin{figure}
\vskip 10mm
 \includegraphics[width=80mm]{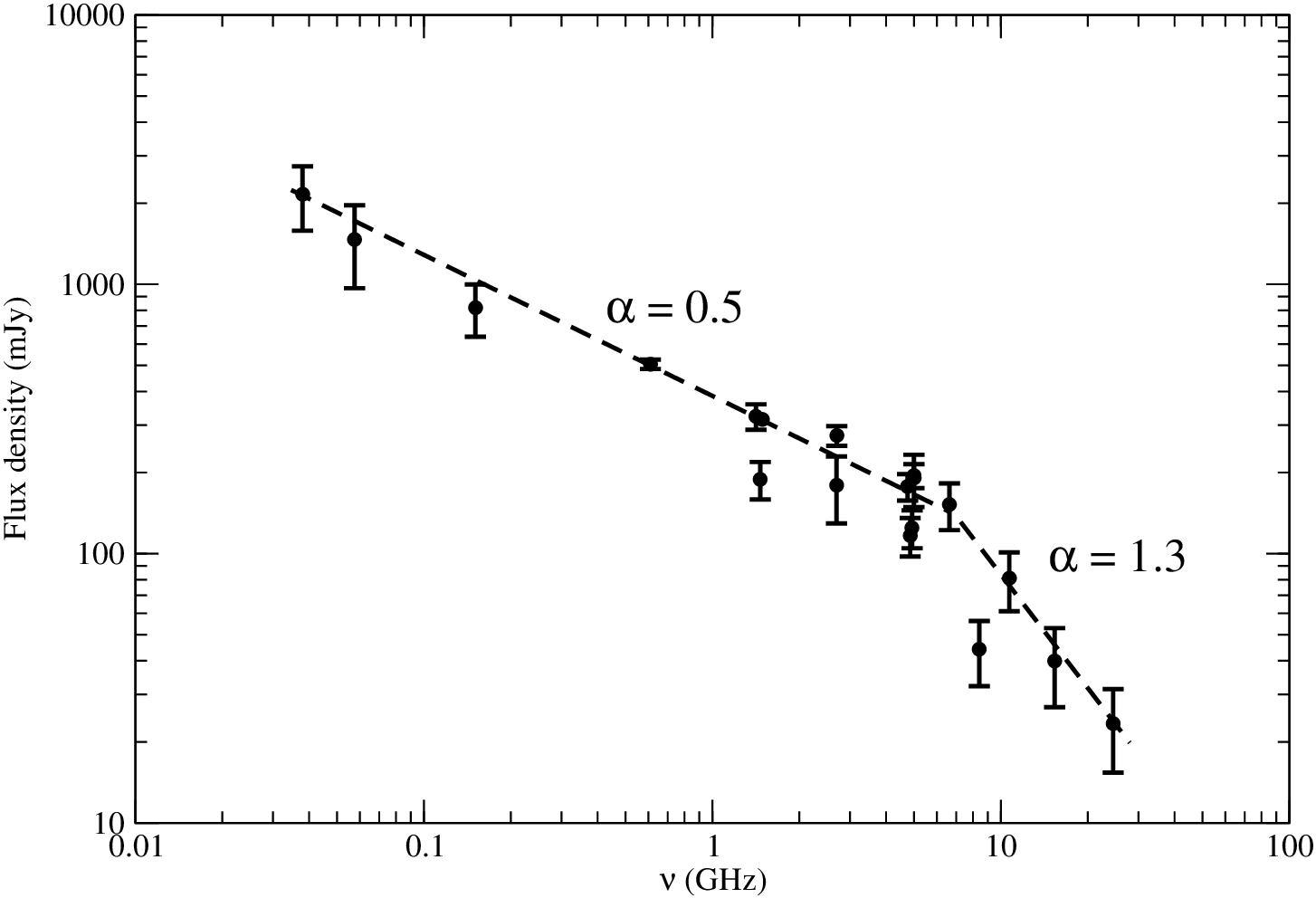}
 \caption{The graph of non-thermal radio flux density $S$ versus frequency $\nu$.}
\vskip 10mm
\end{figure}

\section{Cosmic ray interpretation}
Previous studies show that a simple convective outflow of cosmic ray can account for the radio break \citep{Lisenfeld}. We can estimate the cosmic ray contribution and verify this model. The cosmic ray radio flux depends on $q_{SN}$ (number of electrons or positrons produced per supernova) and $\nu_{SN}$ (supernova rate) \citep{Lisenfeld}. Standard astrophysics predicts $q_{SN} \sim 10^{54}$ \citep{Milne} while $\nu_{SN}$ depends on the total stellar mass $M_*$ and the star formation rate (SFR) \citep{Sullivan}:
\begin{equation}
\nu_{SN}=A \frac{M_*}{10^{10}M_{\odot}}+B \frac{\rm SFR}{M_{\odot}~\rm yr^{-1}},
\end{equation}
where $A \approx (5.3 \pm 1.1)\times 10^{-4}$ yr$^{-1}$ and $B \approx (3.9 \pm 0.7)\times 10^{-4}$ yr$^{-1}$. By taking $M_*=2.8 \times 10^8M_{\odot}$ \citep{Johnson} and the peak $\rm SFR=0.24M_{\odot}~yr^{-1}$ \citep{McQuinn} for NGC 1569, we can get $\nu_{SN} \sim 1 \times 10^{-4}M_{\odot}$ yr$^{-1}$. Since $S \propto q_{SN}\nu_{SN}(E/m_ec^2)^{-\gamma}$ \citep{Lisenfeld}, by taking $\gamma=2$, the cosmic ray contribution is just $S \approx 10$ mJy for $\nu=1.4$ GHz (the non-thermal radio flux density in NGC 1569 $\approx 300$ mJy). In other words, the cosmic ray contribution for the non-thermal radio flux is not very significant. 

Besides, recent analyses based on our galaxy show that cosmic ray transport is mainly driven by magnetic inhomogeneities (the Kolmogorov model) \citep{Recchia,Amato}. The near-disc region is characterized by a Kolmogorov-like diffusion coefficient, but not the outflow velocity. The ion-neutral damping is severe within 1 kpc in our galaxy \citep{Recchia,Amato}. Also, simulations show that gravity is an important factor in convective outflow \citep{Amato}. Therefore, it is dubious that a simple convective outflow model without considering gravity and magnetic inhomogeneities can completely explain the cosmic ray transport in NGC 1569. Furthermore, the high-energy cosmic ray particles would be confined in the galaxy if the cosmic ray transport is mainly driven by magnetic inhomogeneities. In such case, no abrupt break of the spectral index would be found in the radio spectrum. Based on the above reasons, we suspect that dark matter annihilation might be a possible source for the large radio flux of NGC 1569.

\section{Dark matter annihilation model}
Therefore, we investigate whether dark matter annihilation can produce the observed break. We first assume that all the non-thermal radio flux originates from the synchrotron radiation of the electron and positron pairs produced by dark matter annihilation. Since the average magnetic field in NGC 1569 is very high ($B=14 \pm 3\mu$G \citep{Chyzy}), the cooling rate of the electron and positron pairs $b \sim 5 \times 10^{-16}$ GeV s$^{-1}$ is dominated by synchrotron cooling (more than 95\%) \citep{Colafrancesco}. For a 1 GeV electron, the diffusion and cooling timescales are $t_d \sim R^2/D_0 \sim 10^{17}$ s and $t_c \sim 1/b \sim 10^{15}$ s respectively \citep{Colafrancesco}, where we have assumed that the diffusion coefficient is close to the one in Draco dwarf galaxy $D_0 \sim 10^{26}$ cm$^2$ s$^{-1}$ \citep{Colafrancesco2}. Therefore, the diffusion term in the diffusion equation can be neglected and the equilibrium energy spectrum of the electron and positron pairs is proportional to the injection spectrum of dark matter annihilation ($dN_e/dE$) \citep{Storm}. 

Since the diffusion process is not important and the radio emissivity is mainly determined by the peak radio frequency (monochromatic approximation), the total synchrotron radiation flux density (in mJy) of the electron and positron pairs produced by dark matter annihilation at frequency $\nu$ is given by \citep{Bertone,Profumo}:
\begin{equation}
S \approx \frac{1}{4 \pi \nu D^2} \left[ \frac{9 \sqrt{3} (\sigma v)}{2m^2} E(\nu)Y(\nu) \int \rho_{DM}^2dV \right],
\end{equation}
where $D=3.36 \pm 0.20$ Mpc is the distance to NGC 1569, $\rho_{DM}$ is the mass density profile of dark matter, $E(\nu)=13.6(\nu/{\rm GHz})^{1/2}(B/{\rm \mu G})^{-1/2}$ GeV, and $Y(\nu)=\int_{E(\nu)}^m(dN_e/dE')dE'$. By using the Navarro-Frenk-White (NFW) dark matter density profile \citep{Navarro,Schaller} and the data in \citep{Johnson}, we can calculate the total synchrotron radiation flux $S$ within the average radius 2.7' (2.6 kpc) of the galaxy \citep{Lisenfeld}. Since the integral in Eq.~(2) is independent of $\nu$, we can notice that $S \propto \nu^{-1/2}Y(\nu)$. By testing different injection spectrum $dN_e/dE'$ for different dark matter annihilation channels \citep{Cirelli}, we can obtain $S$ as a function of $\nu$ for different annihilation channels and dark matter mass.

To fit the radio data, we first fix the annihilation cross section to be the thermal relic cross section $\sigma v=2.2 \times 10^{-26}$ cm$^3$ s$^{-1}$ \citep{Steigman}. Then we take the dark matter mass $m$ to be a free parameter for each annihilation channel and fit the calculated $S$ with the non-thermal flux data. The one which obtains the smallest reduced $\chi^2$ value ($\chi_{\rm red}^2$) would be the best-fit parameter. Here, we define the reduced $\chi^2$ value as $\chi_{\rm red}^2=(1/f)\sum_i(c_i-o_i)^2/s_i^2$, where $f$ is the degrees of freedom, $c_i$ are the calculated flux, $o_i$ are the observed flux and $s_i$ are the uncertainties of the observed flux. Since there are some discrepancies in the observed radio fluxes near $\nu=1.4-1.5$ GHz, $2.6-2.7$ GHz, $4.7-5$ GHz and $6.6-8.4$ GHz, we calculate the reduced $\chi^2$ value by taking the average of the fluxes for these ranges of frequencies.

In Fig.~1, observational data show $S \propto \nu^{-0.5}$ for small $\nu$. We find that only two possible channels can produce this low-frequency signature: the $e^+e^-$ channel and 4-e channel (annihilation first happens into some new light boson $\Phi$ which then decays into electrons and positrons). It is because only a small number of electron-positron pairs are produced via these two channels for low energy (small $\nu$) so that the function $Y(\nu)$ is nearly a constant (depends very slowly on $\nu$). However, for other channels, $Y(\nu)$ decreases significantly when $\nu$ increases so that the spectral index is somewhat steeper than $0.5$ in the low-frequency regime (e.g. see Fig.~2 for the $b\bar{b}$ quark channel). In the high-frequency regime, nevertheless, the spectral index is not steep enough for the $e^+e^-$ channel to match the observed slope (see Fig.~2). Only the 4-e channel with $m=25$ GeV can give an excellent agreement with the observed spectrum (see Fig.~3). Surprisingly, the best-fit annihilation cross section is exactly the same as the thermal relic annihilation cross section predicted in standard cosmology ($\sigma v=2.2 \times 10^{-26}$ cm$^3$ s$^{-1}$) \citep{Steigman}. The corresponding reduced $\chi^2$ value is $\chi_{\rm red}^2=0.54$.  If we fix the annihilation cross section to be the thermal relic annihilation cross section, the resulting spectrum sensitively depends on the dark matter mass. The reduced $\chi^2$ values change to $\chi_{\rm red}^2=14.4$ and $\chi_{\rm red}^2=7.65$ for $m=20$ GeV and $m=30$ GeV respectively. It is because the corresponding fit deviates very much from an accurate radio flux data point at 610 MHz. 

If we release the annihilation cross section to be a free parameter, good fits can still be obtained for $m=20$ GeV and $m=30$ GeV (see Fig.~4). The annihilation cross sections for $m=20$ GeV and $m=30$ GeV are $1.5 \times 10^{-26}$ cm$^3$ s$^{-1}$ ($\chi_{\rm red}^2=0.72$) and $3.2 \times 10^{-26}$ cm$^3$ s$^{-1}$ ($\chi_{\rm red}^2=0.73$) respectively. Nevertheless, $m=25$ GeV with the thermal relic annihilation cross section via the 4-e channel is still the best model to account for the observed radio spectrum (the smallest $\chi^2$ value). 

We also check our best-fit parameters with the latest gamma-ray observations of MW dSphs galaxies \citep{Ackermann}. Since the 4-e channel is a leptophillic annihilation channel (mainly produce leptons), the number of gamma-ray photons produced is somewhat smaller than that produced from other channels. Based on the observed gamma-ray upper limit, the upper limit of the annihilation cross section for the 4-e channel with $m=25$ GeV is $\sigma v \le 4.5 \times 10^{-26}$ cm$^3$ s$^{-1}$. Therefore, our best-fit annihilation cross section satisfies the most stringent gamma-ray bound. Furthermore, by using our best-fit parameters, the calculated gamma-ray flux produced by annihilating dark matter within $1^{\circ}$ of our galaxy is $\sim 10^{-9}$ cm$^{-2}$ s$^{-1}$, which is just 1\% of the observed flux \citep{Daylan,Ajello}. Therefore, our result is consistent with the latest pulsar interpretation of the GeV excess \citep{Bartels,Ajello}. 

For the AMS constraints, the earlier constraints for the $e^+e^-$ channel is $m \le 90$ GeV for the thermal relic cross section \citep{Bergstrom}. Later, by considering a new phenomenological model, \citet{Cavasonza} obtain a less stringent limit for the $e^+e^-$ channel. Including the systematic uncertainties, the latest constraint is $m \le 50$ GeV (or $\sigma v \le 10^{-26}$ cm$^3$ s$^{-1}$ for $m=25$ GeV) for the $e^+e^-$ channel \citep{Cavasonza}. Since the positron spectrum of the $e^+e^-$ channel is just slightly different from that of the 4-e channel, we assume that the above constraints are also applicable to the 4-e channel. However, the actual picture of the diffusion of high-energy positrons and electrons is very complicated. The most recent analysis in \citet{Cavasonza} follow the parameters and the benchmark diffusion models used in \citet{Cirelli}. In particular, the magnetic field profile assumed follows the one used in \citet{Strong}. This profile is good for large $r$ ($r>1$ kpc) only. The central magnetic field based on this model is just $\sim 10$ $\mu$G while recent studies show that the magnetic field close to $\sim$ mG for the central Milky Way region \citep{Morris}. Therefore, the cooling rate $b$ via synchrotron radiation would be significantly underestimated in \citet{Cavasonza}. Besides, the diffusion coefficient $K$ assumed in the MIN model is approximately a factor of 4 larger than the one obtained in recent studies \citep{Lacroix}. Since the diffusion length scale $\lambda \propto \sqrt{K/b}$ \citep{Lacroix}, a smaller value of $K$ and a larger value of $b$ would give a much smaller value of $\lambda$, which means more high-energy positrons and electrons should be confined within the Milky Way centre. Therefore, the number of the high-energy positrons and electrons calculated in \citet{Cavasonza} is overestimated. In fact, the transport mechanisms of Galactic cosmic rays are still poorly understood. Some recent studies such as \citet{Boudaud} use 1623 transport parameter sets instead of the benchmark models to simulate the diffusion process. The resulting dark matter parameter space increases. Based on the above arguments, the upper limit of the cross section for $m=25$ GeV should be at least a few times larger so that our model is still compatible with the AMS constraints.

\begin{figure}
\vskip 10mm
 \includegraphics[width=80mm]{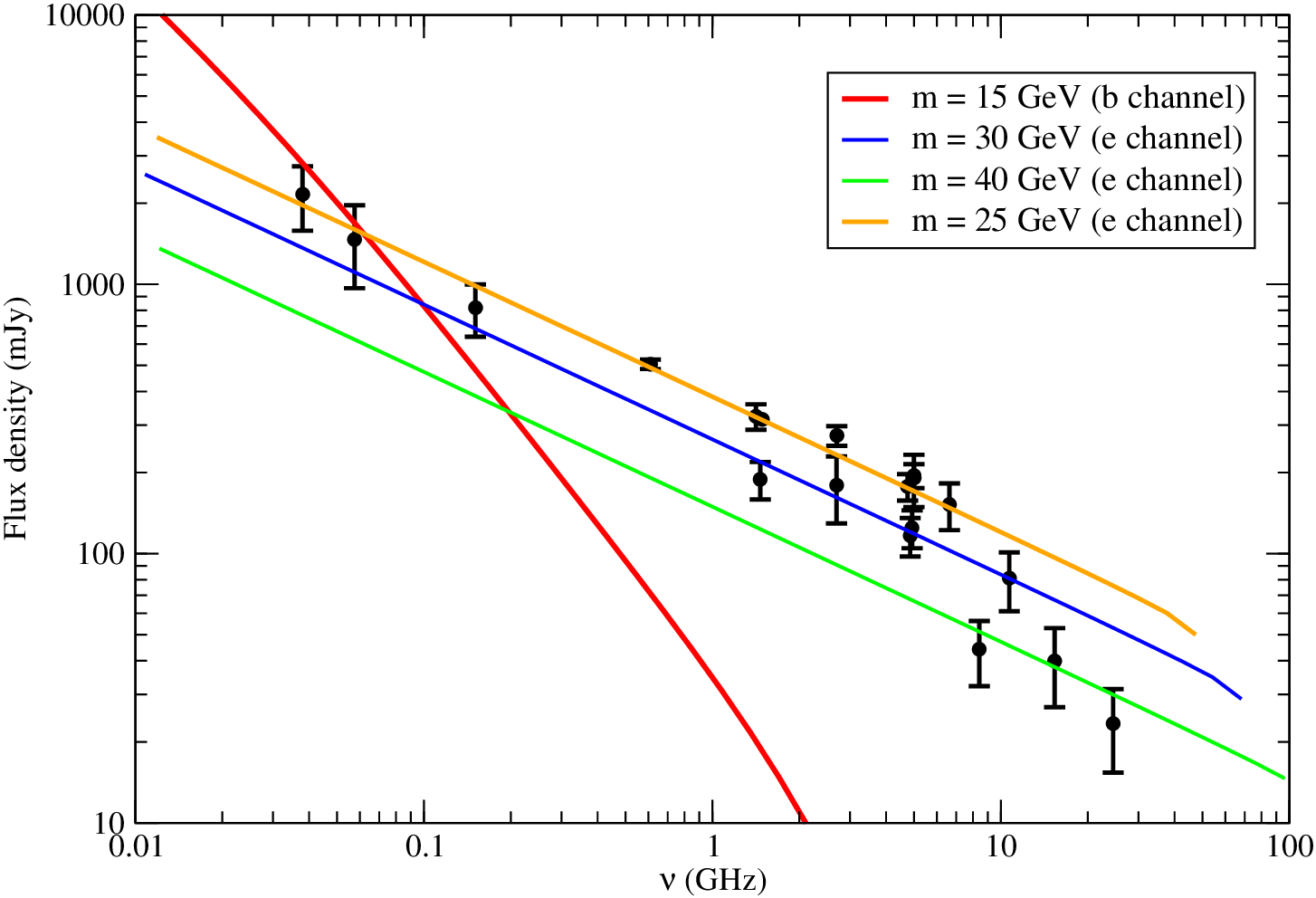}
 \caption{The graph of non-thermal radio flux density $S$ versus frequency $\nu$ for the $e^+e^-$ and $b\bar{b}$ quark annihilation channels. Here, we assume $\sigma v=2.2\times 10^{-26}$ cm$^3$ s$^{-1}$ (the thermal relic annihilation cross section in standard cosmology \citep{Steigman}).}
\vskip 10mm
\end{figure}

\begin{figure}
\vskip 10mm
 \includegraphics[width=80mm]{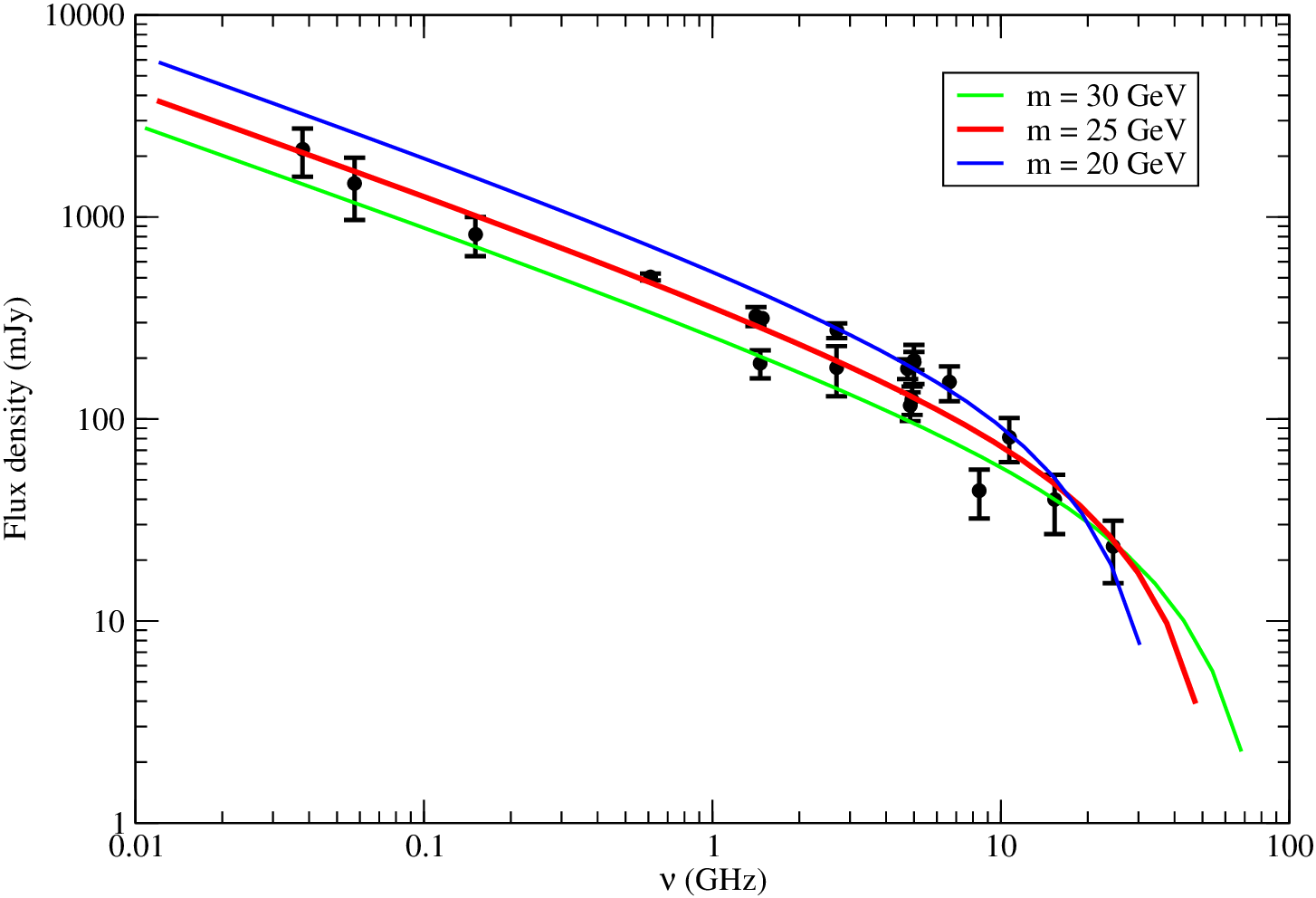}
 \caption{The graph of non-thermal radio flux density $S$ versus frequency $\nu$ for the 4-e annihilation channel. Here, we assume $\sigma v=2.2\times 10^{-26}$ cm$^3$ s$^{-1}$ (the thermal relic annihilation cross section in standard cosmology \citep{Steigman}).}
\vskip 10mm
\end{figure}

\begin{figure}
\vskip 10mm
 \includegraphics[width=80mm]{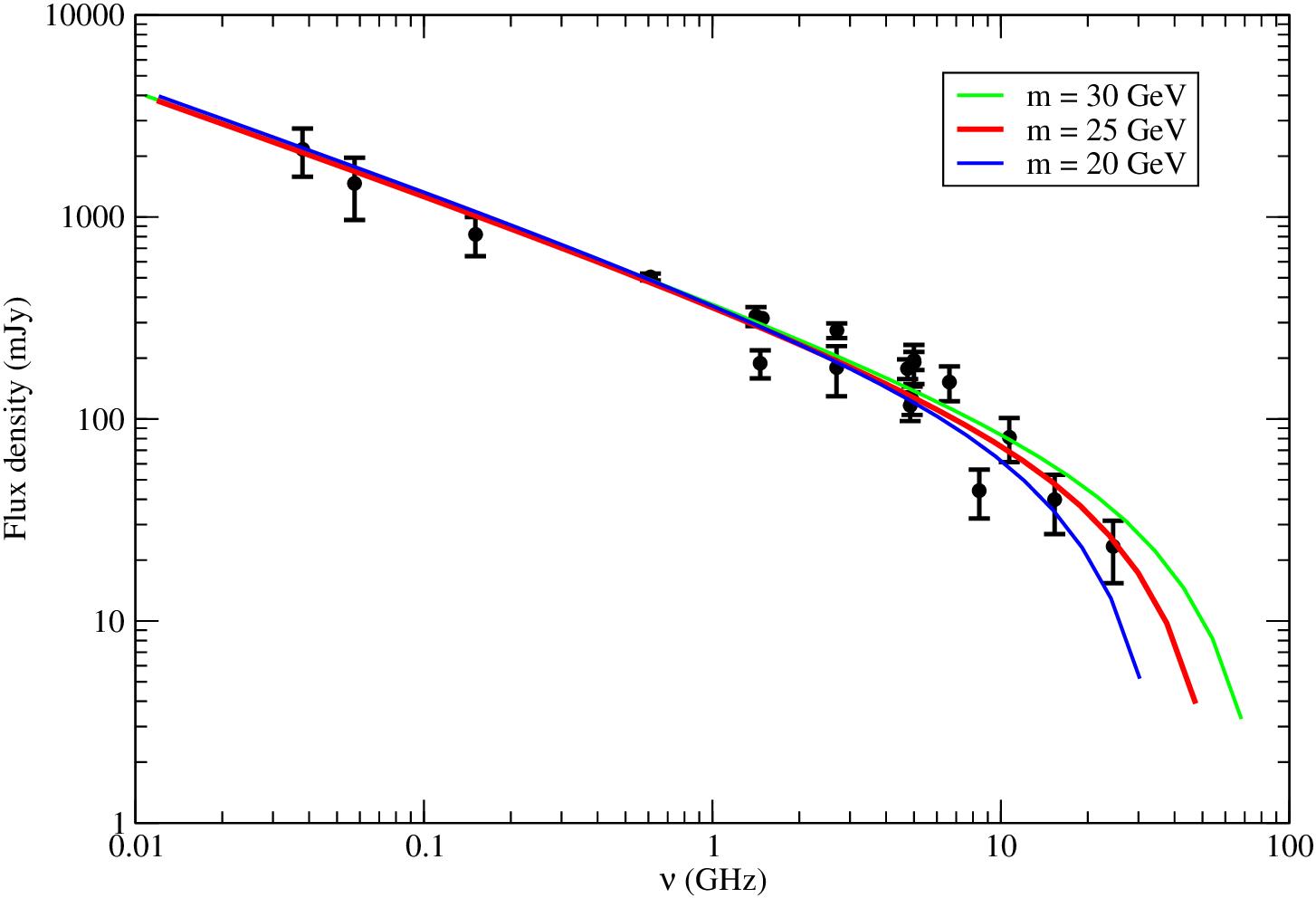}
 \caption{The graph of non-thermal radio flux density $S$ versus frequency $\nu$ for the 4-e annihilation channel. The best-fit annihilation cross sections for $m=20$ GeV, $m=25$ GeV and $m=30$ GeV are $1.5 \times 10^{-26}$ cm$^3$ s$^{-1}$, $2.2 \times 10^{-26}$ cm$^3$ s$^{-1}$ and $3.2 \times 10^{-26}$ cm$^3$ s$^{-1}$ respectively.}
\vskip 10mm
\end{figure}

\section{Discussion}
In this article, we revisit the radio continuum data of NGC 1569 and find that the cosmic ray contribution is not enough to account for the strong radio flux. We propose that the dark matter annihilation model can account for the observed radio flux density. Our analysis shows that dark matter annihilating via the 4-e channel gives an excellent agreement with the observed radio spectrum of NGC 1569. Surprisingly, the best-fit annihilation cross section is equal to the thermal relic annihilation cross section predicted by standard cosmology. Therefore, the observed `radio excess' and the abrupt break in the radio continuum spectrum of a dark-matter-dominated galaxy can be viewed as a unique and promising signature of dark matter annihilation. We predict that dark matter mass is about 25 GeV and there exists an unknown light boson $\Phi$. Further radio observations of other dwarf galaxies and particle search in Large-Hadron-Collider can verify our claim.

\section{acknowledgements}
This work is supported by a grant from The Education University of Hong Kong (activity code: 04256).

\label{lastpage}

\end{document}